\documentclass[10pt]{article} 

\usepackage{amsfonts,amssymb,amsmath}
\usepackage{amsthm,amscd}

\theoremstyle{plain}

\theoremstyle{definition}

\theoremstyle{remark}

\newcommand{\be}{\begin{eqnarray}}
\newcommand{\ee}{\end{eqnarray}}

\renewcommand{\O}{\mathrm{O}}   

\newcommand{\Aut}{\operatorname{Aut}}

\newcommand{\Cset}{\mathbb{C}}

\newcommand{\eq}{{\,  := \, }}

\newcommand{\Hom}{\operatorname{Hom}}
\newcommand{\Hset}{\mathbb{H}}

\newcommand{{\newP}}{{p}}
\newcommand{\newQ}{\lambda}
\newcommand{\neww}{{\omega}}
\newcommand{\newW}{{\varpi}}

\newcommand{\Rset}{\mathbb{R}}
\newcommand{\SL}{\operatorname{SL}}

\newcommand{\SO}{\operatorname{SO}}
\newcommand{\Sp}{\operatorname{Sp}}
\newcommand{\Sq}{\operatorname{Sq}}

\newcommand{\Spin}{\operatorname{Spin}}

\newcommand{\U}{\operatorname{U}}

\newcommand{\unit}{\mathbf{1}}

\newcommand{\Zset}{\mathbb{Z}}

\begin{document}

\begin{titlepage}
\begin{flushleft}
\hfill Imperial/TP/06/JK/01 \\
\hfill {\tt hep-th/0604081} \\
\hfill \\
\hfill 
\end{flushleft}

\vspace*{30mm}

\begin{center}
{\bf \Large Global Spinors and Orientable Five-Branes}\\
\vspace*{8mm}

{Jussi Kalkkinen} \\

\vspace*{3mm}

{\em The Blackett Laboratory, Imperial College} \\
{\em Prince Consort Road, London SW7 2BZ, U.K.} \\

\vspace*{6mm}

\end{center}

\begin{abstract}
Fermion fields on an M-theory five-brane carry a representation of
the double cover of the structure group of the normal bundle. It
is shown that, on an arbitrary oriented  Lorentzian six-manifold,
there is always an $\Sp_{2}$ twist that allows such spinors to be
defined globally. The vanishing of the arising potential
obstructions does not depend on spin structure in the bulk, nor
does the six-manifold need to be spin or spin$^{\Cset}$. Lifting
the tangent bundle to such a generalised spin bundle requires
picking a generalised spin structure in terms of certain elements
in the integral and modulo-two cohomology of the five-brane 
world-volume  in degrees four and five, respectively.

\vfill

\begin{tabbing}
{\em PACS} \hspace{5mm} \= 11.25.-w 
{\em Email}    \>  {\tt j.kalkkinen@imperial.ac.uk}
\end{tabbing}

\end{abstract}
\end{titlepage}



\section{Introduction}
\label{Intro}

The purpose of this paper is to  verify that there are no
obstructions for defining global Fermion fields on an orientable
Lorentzian six-manifold, when the Fermions carry $\Sp_{2}$ charge.
This Fermion structure arises in the chiral $N=2$ tensor theory in
six dimensions \cite{Howe:1983fr}. Though this is the world-volume
theory on an M-theory five-brane
\cite{Duff:1995wd,Witten:1996md,Howe:1997fb}, the analysis
presented here does not make use of assumptions relating to how or
even whether the theory occurs on a brane embedded in an ambient
Supergravity space-time \cite{Cremmer:1978km}.

The spin group $\Spin(1,5)$ can be thought of as  an extension of
the the frame group $\SO(1,5)$ by  $\Zset_{2}$. The fact that the
Fermions  in the present case carry $\Sp_{2}$ charge, leads to extensions of the frame group $\SO(1,5)$ by $\Sp_{2}$,
rather than the usual $\Zset_{2}$. Though such extensions can
contain both $\Sp_{2}$ and $\Spin(1,5)$ as subgroups, this is a
rather more general construction \cite{Avis:1979de}.

The detailed investigation of the actual  lifting procedure allows
one to find which choices precisely need to be made in order to
specify a generalised spin structure. It turns out that there is latitude on two different levels: different ways of finding a lift correspond to a pair of elements, one from $H^{4}(M, \Zset)$ and the other from $H^{5}(M, \Zset_{2})$, where $M$ is the six-manifold under study. Such a pair of elements can be thought of as (a model of) generalised spin structure. An element chosen from the former group has an interpretation in terms of a characteristic class related to  the normal bundle, whereas an element from the latter group has no immediate such meaning.

There are  several reasons to study the five-brane world-volume
theory on its own right as a chiral $N=2$ tensor theory in six
dimensions \cite{Howe:1983fr}. For instance, in a situation where
the brane configuration is not a  stable solitonic solution, the
way the brane is embedded in a geometric bulk space (if
applicable) could be subject to significant dynamical corrections,
as the quantum effects on transverse scalars on a brane could
correct semiclassical brane geometry. In a certain sense, branes
have the capacity to reconstruct their own transverse space. This
is particularly relevant in the study of unstable brane
configurations.

The relationship between the five-brane world-volume theory to the
Supergravity Theory in the bulk is subtle. Verifying that local
anomalies cancel on the five-brane seems to require an 
anomaly inflow mechanism, where the precise way in which the
five-brane is embedded in the ambient space-time plays a pivotal
r\^{o}le \cite{Freed:1998tg}. However, the actual process that
guarantees this cancellation involves cutting quite concretely the
brane off the background. Nevertheless, in a certain large $N$
limit, the world-volume theory seems to lead to an equivalence
beyween the two theories, as the superconformal theory based on
the chiral $N=2$ tensor theory in six dimensions provides a
Holographic dual to Supergravity on $AdS_7 \times S^4$, as
conjectured in \cite{Maldacena:1997re}.

All of this leaves much space for clarification  as to the precise
relationship of the five-brane world-volume to the 11-dimensional
background. Especially unstable backgrounds in mind, it is
important  to divorce world-volume dynamics from bulk phenomena,
as the former could then be used to probe quantum mechanical
backgrounds that are more general than the smooth orientable
spin-manifolds that usually appear in Supergravity backgrounds.

The structure of  generalised spin groups is of course interesting
on its on right, and the present analysis should provide a useful
case study.

The paper is organised as  follows: In Section \ref{Five} the form
of the extension $\Spin_{G}(1,5)$ that the paper concentrates on
is specified. This is the only piece of information from the bulk
that is used, and is forced on one already by the local
world-volume theory. In Section \ref{Chara} the integral and
modulo-two cohomology of the classifying space of generalised spin
structures is found. The general lifting procedure is outlined in
terms of Postnikov-Moore systems in Section \ref{Lift}; this
structure is used in Section \ref{Obs} to show the absence of
obstructions, and to find the cohomology groups where generalised
spin structures are classified. Though the preceding analysis does
not depend on the compactness of the underlying six-manifold, the
vanishing of obstructions and the classification of spin
structures in different non-compact cases is commented on in
Section \ref{Non}. In the final section, Section \ref{Disc},
interpretation of the data required for a generalised spin
structure, consequences of the brane being smoothly embedded in a
spin manifold, and open questions are considered.


\section{Five-brane structure}
\label{Five}

In this section the topological constraints on a stable supersymmetric five-brane, embedded in an 11-dimensional background are reviewed. This sheds light on the local Fermion structure as well. 

The 11-dimensional Supergravity theory \cite{Cremmer:1978km} can
be formulated on orientable Lorentzian manifolds $X$ that have a
spin structure \cite{Witten:1996md}. Orientability is required for
a Lagrangian formulation; spin structure is required for defining
global gravitino fields. Topologically this means that the first
two Stiefel-Whitney classes of the tangent bundle of that manifold
must vanish
\be
w_1(TX) =  w_2(TX) &=& 0 ~. \label{br}
\ee

In this geometric sense a five-brane  --- as a stable,
supersymmetric solitonic solution of Supergravity --- can be
thought of as a smooth Lorentzian submanifold $\iota : M
\hookrightarrow X$ embedded in the 11-dimensional background space
$X$. This means that the tangent and the normal bundles are
related as
\be
\iota^{*}TX &=& TM \oplus NM ~,
\ee
and the above restrictions (\ref{br}) on the bulk space-time imply
\be
w_1(NM) &=& w_1(TM) \\
w_2(NM) &=& w_2(TM) + w_1(TM)^2 ~.
\ee
For a Lagrangian formulation, one assumes again that the five-brane
is orientable
\be
w_1(TM) =  w_1(NM) &=& 0 ~.
\ee

Defining a world-volume theory on a chiral five-brane is subtle
\cite{Witten:1996hc},  first of all as the theory involves a
self-dual three-form field strength \cite{Howe:1983fr}. Apart from
this tensor field, the world-volume supports also scalar fields
and spinorial fields, both of which have to exists as global
fields on the brane.

Spinorial structures enter the {\em local} theory on the
five-brane in the following way \cite{Howe:1983fr}:
\begin{itemize}
\item[-]
The transverse scalars transform, by definition, as a fundamental
$\underline{\mathbf 5}$ of the transverse $\SO(5)$. In the
supersymmetric theory on the brane, however, they show up in the
antisymmetric $\underline{\mathbf 5}$ of $\Sp_2$.
\item[-]
The chiral $\Spin(1,5)$-spinors belong to the fundamental
$\underline{\mathbf 4}$ of $\Sp_2$ subject to a Majorana
condition, as required by $N = (0,2)$ supersymmetry.
\end{itemize}
In this paper defining spinors globally  on the five-brane is
investigated. We shall consider the brane world-volume as an $N=2$
chiral tensor theory, without reference to how or whether it is
embedded in an background space.

If the five-brane should be a spin-manifold $w_{2}(TM)=0$, then
the spinor bundles of the normal and tangent bundles, $S(NM)$ and
$S(TM)$, exist, and world-volume Fermions can be defined globally
as sections of the tensor product bundle $S(TM) \otimes S(NM)$. If
this is not the case and $w_{2}(TM)\neq 0$, a straightforward
spin$^{\Cset}$ structure is of no use, as world-volume
supersymmetry does not allow the Fermions to be electrically
charged with respect to a $\U(1)$ gauge field. Indeed, there is no
Abelian one-form in the massless spectrum of the theory. Instead
of this, the Fermions are coupled to the spin group of the normal
bundle $\Sp_2$,  and one is faced with a somewhat more general
phenomenon.

Given the embedding $\iota : M \hookrightarrow X$, it should be possible to lift the normal and the tangent bundles $NM \oplus TM$ to a spinor bundle together $S(NM \oplus TM)$, even though this bundle might not factorise  in the form $S(TM) \otimes S(NM)$. The reason for this is the fact that the 11-dimensional bundle does exist, and one can always consider its pull-back bundle $\iota^{*}S(TX)$. Note, however, that the present paper does not make use of these arguments  but, for reasons outlined in Section \ref{Intro}, relies entirely on the intrinsic world-volume structure of the brane.

One way to see the effects of this is by noticing that on the
M5-brane the 11-dimensional spin group $\Spin(1,10)$ is broken to
$\Sp_2 \ltimes \Spin(1,5)$. This is a quotient of the direct
product group by the equivalence $(g \cdot \tilde{a}, ~ \alpha )
\sim (g, ~ a \cdot \alpha)$ where $\tilde{a}$ and $a$ generate a
$\Zset_2$ subgroup of the centre of each factor. This group is an
example of generalised spin groups $\Spin_{G}(1,5)$ that fit in
the exact sequence
\be
1 \longrightarrow G \longrightarrow \Spin_{G}(1,5) \longrightarrow
\SO(1,5) \longrightarrow 1   \label{extension} ~.
\ee
It is useful to think of $\Spin_{G}(1,5)$ abstractly as an
extension of $\SO(1,5)$ by $G$, in the same way as the standard
spin group is an extension of the special orthogonal group by
$\Zset_2$. This paper concentrates on the specific extension
that arises on an M5-brane: in the following
\be
\Spin_{G}(1,5) &=& \Sp_2 \ltimes \Spin(1,5) ~,
\ee
where  $G=\Sp_2$, is assumed throughout. Apart from this form of
the generalised spin structure, no other information of the bulk
theory is used in the calculation.


\section{Characteristic classes}
\label{Chara}

Recall that principal bundles $P \longrightarrow M$ with structure group $H$ on a manifold $M$ are in a one-to-one correspondence with the homotopy classes of mappings $f$ from the manifold $M$ to the
corresponding classifying space $BH$. These mappings form the
mapping class group $[M, BH]$. The bundle $P$ is the pull back
$f^*EH$ of the universal $H$-bundle $EH \longrightarrow BH$.
Generators of the cohomology of the classifying space $\omega \in
H^*(BH)$ pull back to characteristic classes $f^{*}\omega$ of the
bundle $P=f^{*}EH$. It is therefore of interest to determine the
cohomology of $B\Spin_G(1,5)$.

The group $\SO(5)$ has one connected component, and $\Spin(5) =
\Sp_2$ is its simply connected compact double cover. The modulo-two and
integral cohomologies of the orthogonal group are
\be
H^*(B\SO(5), \Zset_2) &=& \Zset_2[\bar{w}_2, \bar{w}_3, \bar{w}_4, \ldots] \\
H^*(B\SO(5), \Zset) &=&   \Zset[\bar{\newP}_1, \bar{\newP}_2,
\bar{W}_3, \bar{W}_5, \bar{W}'_7]/\sim ~,
\ee
where the following equivalence should hold in the integral
cohomology
\be
({\bar{W}'_7})^2 &\sim& (\bar{W}_3)^2 \bar{\newP}_2 +
(\bar{W}_5)^2 \bar{\newP}_1 ~. \label{equiva}
\ee

Though the integral cohomology of spin groups is in general
complicated \cite{Quillen}, the  isomorphism $\Spin(5) = \Sp_2$
implies the structure
\be
H^*(B\Sp_{2}, \Zset_2) &=& \Zset_2[\bar{w}_4, \bar{w}_8] \\
H^*(B\Sp_2, \Zset) &=& \Zset[\bar{\newQ}_1,\bar{\newQ}_2]
\ee
as freely generated polynomial algebrae. The generators
$\bar{w}_{4i}$ are modulo-two reductions of the integral
generators $\bar{\newQ}_i$ \cite{Lawson,Thomas}.

Given the fibration \cite{Borel}
\be
B\Zset_2 \hookrightarrow B\Sp_2  \stackrel{\pi}{\longrightarrow}
B\SO(5) ~, \label{pullPI}
\ee
the following relations hold between generators of the
cohomologies of $B\SO(5)$ and $B\Sp_2 $: first of all, the
notation is well-defined in the sense that the generators
$\bar{w}_i$ of the latter are really pull-backs $\pi^*\bar{w}_i$
from the former; secondly, \cite{Thomas}
\be
\pi^*\bar{\newP}_1 &=& 2 \bar{\newQ}_1 \\
\pi^*\bar{\newP}_2 &=& 2 \bar{\newQ}_2 + \bar{\newQ}_1^2 ~.
\ee
The torsion classes in the integral cohomology are related to
modulo-two generators by
\be
\bar{W}_i &=& \beta(\bar{w}_i) \\
\bar{W}'_7 &=& \beta(\bar{w}_2\bar{w}_4) ~.
\ee
The mapping $\beta$ is the Bockstein of the modulo-two short exact
sequence
\be
1 \longrightarrow \Zset \stackrel{2}{\longrightarrow} \Zset
\stackrel{r}{\longrightarrow} \Zset_2 \longrightarrow 1 ~.
\ee
If one denotes by $r$ the reduction of integral classes modulo two, the following results \cite{Brown} hold:
\be
r(\bar{\newQ}_i) &=& \bar{w}_{2i}^2 \\
r \circ \beta &=& \Sq^1 ~.
\ee

The group $\SO(1,5)$ has two connected components. The one
connected to unity fits in the fibration $\Rset^5 \hookrightarrow
\SO^0(1,5) \longrightarrow \SO(5)$. It is therefore contractible
to $\SO(5)$. The maximal compact subgroup of $\SO(1,5)$ is
$\mathrm{S}(\O(1) \times \O(5)) = \Zset_2 \times \SO(5)$. A group
shares its classifying space with its maximal compact subgroup
\cite{Lawson}, so that
\be
B\SO(1,5) &=& B(\Zset_2 \times \SO(5)) = B\Zset_2 \times B\SO(5)~.
\ee
Given the Stiefel-Whitney classes $H^*(B\SO, \Zset_2) =
\Zset_2[w_2, w_3, w_4, \ldots]$ and the results in integral
cohomology of Ref.~\cite{Brown}, one obtains
\be
H^*(B\SO(1,5), \Zset_2) &=&  \Zset_2[\omega] \otimes \Zset_2[w_2,
w_3, w_4, w_5] \\
H^*(B\SO(1,5), \Zset) &=&  \Zset[\varpi] \otimes \Zset[{\newP}_1,
{\newP}_2, W_3, W_5, W'_7]/\sim ~. \label{IntSO}
\ee
The algebra generated in (\ref{IntSO}) is not free, but there is
an  equivalence similar to (\ref{equiva}). The generator $\omega
\in H^*(B\Zset_2,\Zset_2)$ is of degree one and $\varpi \in
H^*(B\Zset_2,\Zset)$ of degree two such that $2\varpi=0$. Note
that on a non-compact Lorentzian manifold, where more specifically the time direction is non-compact, the characteristic classes corresponding to pull-backs of these two generators are trivial.

The connected part of the world-volume spin group is $\Spin^0(1,5)
= \SL(2,\Hset)$. The full group has two components:  one is
connected to $\unit$, the other is connected to the chirality
operator $\chi$. The maximal compact subgroup is $\Sp_2$, so
\be
B\Spin(1,5) &=& B\Zset_2 \times B\Sp_2 ~, \label{Bspin}
\ee
This implies
\be
H^*(B\Spin(1,5), \Zset_2) &=&  \Zset_2[\omega]  \otimes \Zset_2[w_4, w_8] \label{spin1}\\
H^*(B\Spin(1,5), \Zset) &=&  \Zset[\varpi] \otimes
\Zset[{\newQ}_1, {\newQ}_2] ~, \label{spin2}
\ee
with ${w}_{4i} = r({\newQ}_i)$.

Spinors on an M-theory five-brane carry both a representation of
$\Sp_{2}$ and $\Spin(1,5)$. This means that the {\em physical}
extension $\Spin_G(1,5)$ is of the form
\be
\Zset_2 \hookrightarrow \Sp_2 \times \Spin(1,5) \longrightarrow
\Spin_{G}(1,5) ~,
\ee
where the image of nontrivial element $-\unit$ in the fibre is $(-\unit, -\unit)$ in the total space. This leads to the fibration
\be
B\Zset_2 \hookrightarrow B\Sp_2 \times B\Spin(1,5) \longrightarrow
B\Spin_{G}(1,5) ~. \label{fibB}
\ee
Note that the $\Zset_2$ factor that appears in (\ref{Bspin}) has a
nontrivial image on the base. This means that the fibration
(\ref{fibB}) is nontrivial. Using the standard Leray-Serre
spectral sequence, one finds the cohomology of the base space to
be
\be
\lefteqn{H^*(B\Spin_G(1,5), \Zset_2) } \nonumber \\
&& = \Zset_2[\omega] \otimes \Zset_2[{w}_4,\bar{w}_4, {w}_8, \bar{w}_8] \otimes
\Zset_2[{\neww}_2, {\neww}_3, {\neww}_5, {\neww}_{9}, \ldots ]  \label{cSG1} \\
\lefteqn{H^*(B\Spin_G(1,5), \Zset) } \nonumber \\
&& = \Zset[\varpi] \otimes \Zset[{\newQ}_1,\bar{\newQ}_1,
{\newQ}_2,\bar{\newQ}_2] \otimes \Zset[{\newW}_3, {\newW}_5,
{\newW}_{9}, {\newW}_{17},  \ldots] \label{cSG2} ~,
\ee
where in the last factor only generators ${\neww}_i$ (resp.~${\newW}_i$) with degree of the form $i=2^{r+1}$ appear.

This structure arises because (\ref{fibB}) is a nontrivial
fibration, and as such the transgression $d_{2}: H^{1}(B\Zset_{2},
\Zset_{2}) \longrightarrow H^{2}(B\Spin_{G}(1,5),\Zset_{2})$ is
nontrivial. On the level of the Leray-Serre spectral sequence this
means first of all that the derivative  $d_2$ acting on
$E^{*,*}_2$ has to be nontrivial $d_2(a) = {\neww}_2 \neq 0$
where $a \in H^1(B\Zset_2,\Zset_2) $ is the generator. By
Corollary 6.9 of \cite{McCleary} this leads to a tower of
generators
\be
d_{ 2i+2}(a^{2^{i+1}}) =  d_{ 2i+2}(\Sq^i a^{2^{i}})= \Sq^i
d_{i+1} a^{2^{i}} = \Sq^i {\neww}_{2^i +1} = {\neww}_{2^{i+1} +1}
~.
\ee
Though Corollary 6.9 of \cite{McCleary} does not apply  over
integers, a similar argument can be devised, and the above result
follows.

The spectral sequence keeps track of the various cohomology groups not as rings but simply as graded vector spaces. Therefore, the generators $\neww_i$ and $\newW_i$ have, a priori, nothing to do with Stiefel-Whitney classes. The generators that are a priori  related to Stiefel-Whitney classes appear in (\ref{spin1}) and (\ref{spin2}). The analysis of Section \ref{Lift} is required to clarify this issue, and it turns out that the new generators ${\neww}_i$ should a posteriori be identified with universal Stiefel-Whitney classes, at least up to degree five.

The fact that the cohomology involves an infinite tower  of
generators should not be surprising. First of all, Spin groups do
have higher degree generators than SO groups; an example is the
degree eight class of $\Spin(5)=\Sp_{2}$ whereas the generators of
characteristic classes of $\SO(5)$ go only up to degree five. It
turns in fact out (\cite{Borel2}, Prop.~15.2) that $H^{*}(B
\Spin(n),\Zset_{2})$ is a polynomial algebra precisely for $n \leq
9$. As $\Spin_{G}(1,5)$ is in a certain sense a reduced form of
$\Spin(1,10)$, one should perhaps expect to find such a rich
structure.

Section \ref{Obs} provides a partial consistency check to these
results.


\section{Lifting}
\label{Lift}

The exact sequence (\ref{extension}) induces the fibration
\cite{Borel} of classifying spaces
\be
BG \hookrightarrow B\Spin_{G}(1,5)  \longrightarrow B\SO(1,5) ~. \label{fibu}
\ee
In what follows, the mapping class group $[M,B
\Spin_{G}(1,5)]$ is analysed in terms of the Moore-Postnikov system of this fibration. {These techniques were first introduced in the Physics
Literature in \cite{Avis:1979de}.}

The Moore-Postnikov system of the fibration (\ref{fibu}) of the
classifying space  $E \eq  B\Spin_{G}(1,5)$ consists of a sequence of fibrations
\be
p_{n}: E^{[n]} \longrightarrow E^{[n-1]} \label{moorepostnikov1}
\ee
with fibre $K(\pi_{n}(BG),n) = K(\pi_{n-1},n)$. At level $n=0$ the space is simply the base space $ E^{[1]} = B\SO(1,5)$; for each $n>1$, $E^{[n]}$ has the
homotopy groups $\pi_i(E)$ for $i\leq n$, and $0$ otherwise. The
larger the index $n$, the better approximation $E^{[n]}$ is of the
total space $E$.  Each fibration can be chosen in terms
of the Postnikov invariants
\be
\Big[ k^{n+1} : E^{[n]} \longrightarrow \hat{K}(\pi_{n-1}, n+1) \Big]
\label{inv1} ~,
\ee
which are homotopy classes of mappings. Using the isomorphism
\be
[M, K(\pi,n)] &\simeq& H^n(M,\pi) ~, \label{iso}
\ee
these invariants can be considered elements of $H^{n+1}(E^{[n]},
\pi_{n-1})$.

In a non-simply connected case  \cite{Robinson} the
Eilenberg-MacLane spaces $\hat{K}(\pi_{i},n)$ appearing in the
Postnikov invariants are certain twisted versions of the
standard spaces ${K}(\pi_{i},n)$. In $H^n(M,
\tilde\pi_i)$ the coefficient sheaf $\tilde\pi_i$ is the constant
sheaf $\pi_i$  twisted by the action of an element of $\Aut\pi_i$
over noncontractible paths in the total classifying space $E$.

In the present case there is precisely  one such non-contractible
path, as the total classifying space has the fundamental group
$\pi_1(E) = \pi_0(\Spin_{G}(1,5)) = \Zset_2$. Along these paths a
tangent vector picks up a holonomy from the component of
$\SO(1,5)$ connected to a total reflection $-\unit \in SO(1,5)$.
On the level of the spin group, this rotation lifts to $\pm\chi
\in \Spin(1,5)$, the chirality operator.

On a Lorentzian manifold $M$ the tangent bundle splits to an $\O(1)
\times \O(n)$ bundle $TM \simeq V^{-} \oplus V^{+}$. In the
orientable case, the obstruction $w_{1}(TM) = w_{1}^{+} +
w_{1}^{-} =0$ relates the two logically distinct obstructions
$w_{1}^{\pm}$ for either one of the vector bundles $V^{\pm}$ to be
separately orientable. We shall refer to $w_{1}^{-}$ as the obstruction to a {\em
temporal orientation}. It has a chance to be nontrivial if the
six-manifold $M$ is compact.

As $B\SO(1,5)$ has the same homotopy type as $E$  up to
four-sceletons, one can replace $E$ here by $B\SO(1,5)$, and the
question of twisting coefficient sheaves $\tilde\Zset =
\Zset_\omega$ reduces to a choice of an element in $\omega \in
H^1(B\SO(1,5), \Zset_2) = \Zset_{2}$. The nontrivial element there
is the obstruction to a temporal orientation $\omega = w_{1}^{-}$. The twisting, if
any, is then done by using the temporal orientation sheaf as the
coefficient sheaf in the cohomologies where the obstructions have
their values.

The relevant part of the Moore-Postnikov system can be
conveniently represented as the diagram
\be
\begin{CD}
K(\pi_{4},5) @>>> E^{[6]} \\
&& @Vp_6VV \\
K(\pi_{3},4) @>>> E^{[5]} @>k_6>> \hat{K}(\pi_{4},6) \\
&& @Vp_5VV \\
&& B\SO(t,s) @>k_5>> \hat{K}(\pi_{3},5) ~.
\end{CD}
\ee
We have denoted here $\pi_i = \pi_i(\Sp_2)$. Here the following facts have been used:
\begin{itemize}
\item For bundles on a six-manifold $M$ it is
sufficient to consider the tower up to six-sceletons $[M, E] =
[M,E^{[6]}]$;
\item Similarly, $[M,K(\pi,n)] = H^{n}(M,\pi)=0$ for $n >6$.
\item The homotopy groups  $\pi_0 = \pi_{1} = \pi_{2} =0$
are trivial in all of the present constructions.
\end{itemize}

Taking the corresponding mapping class groups and putting in place $\pi_3 = \Zset$ and $\pi_4 = \Zset_2$, one finds the two exact sequences
\be
H^{5}(M, \Zset_2) \longrightarrow \Big[M, B\Spin_{G}(1,5)\Big]
&\stackrel{p_{6*}}{\longrightarrow}& \Big[M, E^{[5]}\Big]
\stackrel{k_{6*}}{\longrightarrow} H^{6}(M,  \Zset_2) \nonumber \\
H^{4}(M,  \Zset) \longrightarrow \Big[M, E^{[5]}\Big]
&\stackrel{p_{5*}}{\longrightarrow}& \Big[M, B\SO(1,5) \Big]
\stackrel{k_{5*}}{\longrightarrow} H^{5}(M,  \tilde\Zset ) ~. \nonumber
\ee
There are no nontrivial automorphisms of $\Zset_{2}$, so the twisting is trivial in $H^{6}(M,  \Zset_2)$.


\section{Obstructions}
\label{Obs}

In the two  exact sequences of the last section the cohomology groups on the right are obstructions for lifting the  tangent bundle $TM$,  as represented by a class $[\zeta] \in [M,
B\SO(1,5) ]$,  to an element of $[M, E^{[5]}]$ and then from there to a generalised spin bundle, as represented by a class in $[M,
B\Spin_{G}(1,5)]$. The cohomology groups on the left describe the latitude in the lifting procedure, and can be thought of as classifying generalised spin structures.  

The map $k_5$ determines a class in $[k_5] \in H^5(B\SO(1,5), \Zset)$. Given the mapping $\zeta : M \longrightarrow B\SO(1,5)$ corresponding to the tangent bundle $[\zeta] = TM$, one has
\be
k_{5*}[\zeta] &=& [k_5 \circ \zeta ] = \zeta^*[k_5] ~.
\ee
If this obstruction vanishes, there is a lift of $\zeta$ to $[\hat\zeta] \in [M, E^{[5]}]$ that satisfies $p_{5*}[\hat\zeta]= [\zeta]$.
Similarly, to lift $\hat\zeta$ further to a class $\xi \in [M, E^{[6]}] = [M, B\Spin_{G}(1,5)]$, the obstruction
\be
k_{6*}[\xi] &=&  \xi^*[k_6]
\ee
must vanish.

In Section \ref{Chara} it has been shown that all the generators of the cohomology of the classifying space $B\SO(1,5)$, where one lifts a class $\zeta\in [M, B\SO(1,5)]$ from, are present in the cohomology of the classifying space $B\Spin_{G}(1,5)$, to whose mapping class group one is lifting it. As there are no such missing generators, there should not be universal obstructions, and it should always be possible to choose a lift of $\zeta = TM$ to a $\xi \in [M, B\Spin_{G}(1,5)]$ in such a way that $p_{5*}\circ p_{6*}[\xi]= [\zeta]$. The results of Section \ref{Chara} leave two questions open, however:
\begin{itemize}
\item It needs to be shown that the degree five generator $\omega_{5}$ is indeed the Stiefel-Whitney class $w_{5}$ and not a new generator;
\item The lifting procedure proceeds in two steps: it needs to be shown that there are no missing generators in the cohomology of $E^{[5]}$ that would then somehow be restituted in the cohomology of $E^{[6]}$.
\end{itemize}
In this section it is shown that the identification $\omega_{5} =w_{5}$ is valid. There will indeed turn out to be space for an intermediate obstruction to lifting from $E^{[5]}$ to $E^{[6]}$, but it is shown below that this obstruction vanishes in the present case.

In the process of performing such a lift, one must make independent choices that amount to picking an element first from $H^{4}(M, \Zset)$ and then from $H^{5}(M, \Zset_2)$. In order to investigate this process further, and to have a partial check for the results presented in Section \ref{Chara}, the lifting procedure and the vanishing of the obstructions in the Moore-Postnikov system of Section \ref{Lift} is verified in detail in this section.


\subsection{Integral obstructions}
\label{ObsInt}

By Hurewicz Isomorphism, the first nontrivial integral cohomology
group of $K(\Zset,4)$ is generated by the fundamental class
$I_{4}$
\be
H^{4}(K(\Zset,4),\Zset) &=& \Hom(H_{4}(K(\Zset,4),\Zset), \Zset) \\
&=& \Hom (\Zset,\Zset) = \Zset ~.
\ee
It follows from the Leray-Serre spectral sequence that $E^{[5]}$
has the same cohomology groups as $B\SO(1,5)$ in degrees one, two,
and three. In particular, one should identify
\be
W_3 &=& \newW_3 ~.
\ee

In degree four, the total space $E$ is known to have one more
generator $\bar{\newQ}_{1}$ than the base space $B\SO(1,5)$. As
the only available extra generator is the fundamental class
$I_{4}$ of the fibre, one must identify $\bar{\newQ}_{1} = n I_4$
for some integer $n$. Other multiples $k I_{4}$ for $k<n$ would then have to be eliminated by setting $d_{5}(kI_{4})$ to be nontrivial. This could be done consistently only for $n = 1,2$. In next section one discovers, however, that $\iota_{4} = r(I_{4})$ remains in the cohomology, and at the very latest in studying modulo-two cohomology one discovers $n=1$,
\be
\bar{\newQ}_{1} &=& I_4 ~,
\ee
and $d_{5}(I_{4})=0$.

As none of the higher approximations $E^{[n]}$, $n>5$ will change
integral cohomology in degrees less than five,  one has for $i
\leq 4$
\be
H^{i}(E,\Zset_{2}) &=& F^{i} \Big( H^{*}(B\SO(1,5),\Zset) \otimes \Zset[\bar{\newQ}_{1}] \Big) \\
&=& F^{i} \Big(  \Zset[\varpi, W_3, {\newQ}_1, \bar{\newQ}_{1} ]
\Big) \label{fiveint} ~,
\ee
where $F^i$ filters out the degree $i$ part from the ring. This is
fully consistent with (\ref{cSG2}).

Without knowing the fifth cohomology group of the fibre $H^{5}(K(\Zset,4),\Zset)$ one cannot write down the precise fifth cohomology group  $H^{5}(E^{[5]},\Zset)$. However, as it is known that the differential $d_{5}=0$ is trivial, it is clear that $H^{5}(E^{[5]},\Zset)$ includes all generators of the base $H^{5}(B\SO(1,5),\Zset)$. The next differential $d_{6}$ could eliminate generators from the cohomology of the base, but only at degree six $H^{6}(B\SO(1,5),\Zset)$.

As $H^{5}(E^{[5]},\Zset)$ contains all the classes of $H^{5}(B\SO(1,5),\Zset)$, there is no obstruction to the lift. Choosing such a lift requires making a choice, which amounts to picking the class
\be
\hat\zeta^{*}\bar{\newQ}_{1} &\in& H^{4}(M, \Zset) \label{pick}
\ee
that corresponds to half the first Pontryagin class of the normal
bundle.


\subsection{Modulo-two obstructions}

Up to degree six, the homotopy type of $E = B\Spin_G(1,5)$ is
$E^{[6]}$.   The approximations $E^{[5]}$ and  $E^{[6]}$ fit in
the fibrations
\be
K(\Zset_2,5) &\hookrightarrow& E^{[6]} \longrightarrow E^{[5]}  \label{es1} \\
K(\Zset,4) &\hookrightarrow& E^{[5]} \longrightarrow B\SO(1,5) \label{es2}  ~.
\ee
The modulo-two cohomology groups of the fibres are
\be
H^{*}(K(\Zset_2,5),\Zset_{2}) &=& \Zset_{2}[\Sq^{I}] = \Zset_{2}[\iota_{5}, \Sq^{1} \iota_{5}, \Sq^{2} \iota_{5}, \ldots ] \label{c1} \\
H^{*}(K(\Zset,4),\Zset_{2}) &=& \Zset_{2}[\Sq^{J}] = \Zset_{2}[\iota_{4}, \Sq^{2} \iota_{4},  \Sq^{3} \iota_{4},   \ldots ]~,  \label{c2}
\ee
where $\iota_n$ are the generators, and the multi-indices $I$ and $J$ are appropriately restricted \cite{McCleary}.

As the lowest element $\iota_{4}$ in $H^{*}(K(\Zset,4),\Zset_{2})$
is  of degree four, it follows from the Leray-Serre spectral
sequence that $E$ has the same cohomology groups as $B\SO(1,5)$ in
degrees one, two, and three. In particular, one should  identify
\be
\neww_2 &=& w_2 \\
\neww_3 &=& w_3 ~.
\ee
In degree four, the total space $E$ is known to have one more
generator $\bar{w}_{4}$ than the base space $B\SO(1,5)$. The only
available new generator in that degree is the fundamental class of
the fibre $\iota_{4}$. To keep it in the cohomology, it must be
transitive in (\ref{es2}), that is $d_{r}(i_{4}) = 0$ for $r \geq 2$. We identify
\be
\iota_{4} &=& \bar{w}_{4} ~.
\ee
This is consistent with the fact that the $\Sp_{2}$ class is a reduction of an integral class $\bar{\newQ}_{1}$ as so is $\iota_{4} = r(I_{4})$. Indeed, there is no $\Sq^{1}\iota_{4}$ generator in (\ref{c2}).

There is no new generator in degree five in the fibre; the next new generators are $\Sq^{2}\iota_{4}$ in degree six, and $\Sq^{3}\iota_{4}$ in degree seven. By choosing $d_{7}(\Sq^{2}\iota_{4})$ suitably, one could either keep or eliminate this generator. As there is no generator of degree seven in the cohomology of the base space, however, one expects this differential to be trivial, and $\Sq^{2}\iota_{4}$ to remain in the cohomology; the same applies to the generator $\Sq^{3}\iota_{4}$ in degree seven. We shall see below that this is indeed the only consistent choice in these degrees. Filtering out the degrees already analysed $i \leq 7$, the modulo-two  cohomology of $E^{[5]}$ is therefore
\be
H^{i}(E^{[5]},\Zset_{2}) &=& F^{i} \Big(H^{*}(K(\Zset,4),\Zset_{2})
\otimes H^{*}(B\SO(1,5),\Zset_{2}) \Big) \\
&=& F^{i} \Big(\Zset_{2}[\bar{w}_{4}, \Sq^{2}\bar{w}_{4}, \Sq^{3}\bar{w}_{4}] \otimes
 \Zset_{2}[\omega, w_{2},w_{3},w_{4},w_{5}] \Big)
~.
\ee

The Leray-Serre spectral sequence of fibration (\ref{es1}) implies
now that the cohomology of the total space $E^{[6]}$ and that of $E^{[5]}$ coincide in degrees up to and including four. At degree five there is, potentially, a new generator $\iota_5$, which is the
fundamental class of the fibre $K(\Zset_2,5)$. As known from Section \ref{Chara}, there should not be one, so that $d_{6}(\iota_{5}) \neq 0$ must be a nontrivial element in $H^{6}(E^{[5]},\Zset_{2})$. The free ring structure determines
\be
d_{6}(\iota_{5}) &=& \Sq^{2}\bar{w}_{4} ~. \label{el5}
\ee
In hindsight, leaving $\Sq^{2}\bar{w}_{4}$ in the cohomology of $E^{[5]}$ was, therefore, justified. The choice of $d_{6}$ removes now $\iota_{5}$ from degree five and both $\Sq^{2}\bar{w}_{4}$ and $\omega \iota_{5}$ from degree six.

There is a generator of degree six in the fibre $\Sq^{1}\iota_{5}$, and $d_{6}(\Sq^{1}\iota_{5})=0$. The nontrivial differential (\ref{el5}) implies  also
\be
d_{7}(\Sq^{1}\iota_{5}) &=& \Sq^{1}\Sq^{2}\bar{w}_{4} \\
&=& \Sq^{3}\bar{w}_{4} ~.
\ee
This generator is indeed in the cohomology of $E^{[5]}$, but is now eliminated together with $\Sq^{1}\iota_{5}$ from the cohomology of $E^{[6]}$. We see, then, in hindsight that the generators $\Sq^{2}\bar{w}_{4},\Sq^{3}\bar{w}_{4}$ were indeed both required in the cohomology of $E^{[5]}$ so that the generators $\iota_{5},\Sq^{1}\iota_{5}$ could be eliminated from the cohomology of $E^{[6]}$ consistently with the requirements of Section \ref{Chara}.

The rest of the generators $\Sq^{I}\iota_{5}$ of the cohomology of the fibre $K(\Zset_{2},5)$ have potentially nontrivial transgressions as well
\be
d_{p+6}(\Sq^{I}\iota_{5}) &=& \Sq^{I}\Sq^{2}\iota_{4} ~.
\ee
Here $p$ is the degree of the multi-index $I$. Due to the different structures of the two cohomologies (\ref{c1}) and (\ref{c2}), not all elements in the latter are in the image of $d_{p+6}$; this should account for the tower of generators found in (\ref{cSG1}), and amounts to a   consistency check up to degree six for the results in Section \ref{Chara}.

Therefore, filtering out degrees $i \leq 6$ by $F^{i}$
\be
H^{i}(E^{[6]},\Zset_{2})  &=& F^{i}  \Big(  H^{i}(E^{[5]},\Zset_{2})/_{\Sq^{2}\bar{w}_{4}=\Sq^{3}\bar{w}_{4}=0}  \Big) \\
&=& F^{i} \Big( \Zset_{2}[\omega,w_{2},w_{3},w_{4},\bar{w}_{4},w_{5}] \Big) \label{sixm2}
~.
\ee
The lowest generators inherited from the universal Stiefel-Whitney classes of $B\SO(1,5)$, namely $w_{2},w_{3},w_{4}$, had no chance of getting eliminated in the above procedure. The fact that there was an extra generator $\bar{w}_{4}$ in degree six meant that one could not eliminate the fifth Stiefel-Whitney class $w_{5}$ either. One identifies, then,
\be
\neww_5 &=& w_5 ~.
\ee
It is interesting to note that though $\Sq^{1}\bar{w}_{4} =0$ in the fibre in this construction, a nontrivial class $\Sq^{1}{w}_{4} = w_{5}$ is allowed on the base $B\SO(1,5)$. 

The only characteristic class in (\ref{sixm2}) that is not inherited from the cohomology of $ B\SO(1,5)$ is $\bar{w}_{4}$, the modulo-two generator of the cohomology of the $K(\Zset,4)$-fibre. As seen in Section \ref{ObsInt}  in particular, it is there already in $H^{*}(E^{[5]},\Zset_{2})$.

One is now in a position to comment on the lifting procedure from $[M, E^{[5]}]$ to $[M,E]$. There is precisely one generator in degree six that is present in $E^{[5]}$ but not in $E^{[6]}$ due to (\ref{el5}). This generator is the obstruction, and consistency requires
\be
\hat\zeta^{*} \Sq^{2}\bar{w}_{4} &=& 0 ~.
\ee
Recall that Steenrod squares commute with pull-backs. We can use the $\pi^{*}$ of (\ref{pullPI}) in a given fibre
\be
\Sq^{2}\bar{w}_{4} = \Sq^{2} \pi^{*} \bar{w}_{4} = \pi^{*} \Sq^{2}\bar{w}_{4} = \pi^{*}(\bar{w}_{2}\bar{w}_{4}+\bar{w}_{6}) =0 ~,
\ee
because both $\bar{w}_{2}$ and $\bar{w}_{6}$ pull back to zero in the cohomology of  $B\Sp_{2}$.

There is then no obstruction to lifting an element of $[M, E^{[5]}]$ to $[M,E]$. Such a lift requires making a choice; the latitude in different choices corresponds to elements of $H^{5}(M, \Zset_2)$. Fixing this latitude does not correspond to choosing a characteristic class of $B\Spin_{G}(1,5)$, as opposed to the one performed in (\ref{pick}).


\section{Noncompact cases}
\label{Non}

Thus far it was assumed only that $M \subset X$ is a {\em smooth orientable Lorentzian six-manifold}. For simplicity, in this section only, it is assumed to be also connected; the existence and classification of global spinors can clearly be discussed
component by component if it is not.

In this section, and in this section only, it is further assumed that the five-brane world-volume $M$  is {\em contractible} to some compact manifold $M_{n}$ of dimension $n$. This happens, for instance, when the total world-volume is of the form
\be
\Rset^{6-n} \hookrightarrow M \longrightarrow M_{n}~.
\ee
This simplifies matters, as the homotopy axiom of cohomology ---
which holds also   with local, or twisted, coefficients ---
guarantees  $H^{*}(M) = H^{*}(M_{n})$.  There are four basic
cases:
\begin{itemize}
\item At $n=6$ the world-volume, including the time direction, is compact. To analyse this case the full power of the results of the previous sections are needed.
\item At $n=5$ there is one non-compact direction that one can think of as time. This case is discussed briefly in this section without recourse to the detailed structure of $B\Spin_{G}(1,5)$.
\item At $n=4$ the obstructions vanish obviously, and the lifts are classified in $H^{4}(M,\Zset)=\Zset$.
\item For $n \leq 3$, the obstructions vanish, and there is a canonical lift.
\end{itemize}

Consider a class $\zeta \in [M, B\SO(1,5)]$ in the case $n=5$, so that $M$ is contractible to a five-dimensional compact manifold $W$: then
\be
H^{6}(M,\Zset_{2}) &=& 0 \\
H^{5}(M,\Zset_{2}) &=& \Zset_{2} \\
H^{5}(M,\tilde\Zset) &=& \begin{cases}
\Zset  & \text{trivial twist  } \zeta^{*}\omega = 0 \\
\Zset_{2}  & \text{non-trivial twist} ~,
\end{cases}
\ee
where $\omega \in H^{1}(B\Spin_{G},\Zset_{2})$, and the class $\zeta^{*} \omega \in H^{1}(M, \Zset_{2})$ determines the character of the time orientation sheaf.

If one takes the non-compact direction to be time, then $\zeta^{*}\omega = 0$, and the fifth cohomology is taken over trivially twisted coefficients.

The homomorphism $k_{5*}$ is defined on the fifth cohomology of the classifying space $H^5(B\SO(1,5))$, whose elements are all two-torsion $2 H^5(B\SO(1,5))=0$. The target space, however, has no torsion elements $H^{5}(M,\Zset) = \Zset$. It follows that the mapping must be trivial $k_{5*}=0$. There is then no obstruction to lifting the tangent bundle to a generalised spin bundle when the brane is compact but the time-direction is not.

Note that if the time direction is compact and the cohomology is nontrivially twisted, one must make recourse to the arguments in the rest of the paper to show the vanishing of the obstruction.


\section{Discussion}
\label{Disc}

We have shown that on an oriented Lorentzian five-brane world-volume there is no obstruction for lifting the tangent bundle to a generalised spin bundle with structure group $\Spin_{G}(1,5)=\Sp_2 \ltimes \Spin(1,5)$. In the process of eliminating obstructions, the cohomology of the classifying space of generalised spin bundles was found
\be
\lefteqn{H^*(B\Spin_G(1,5), \Zset_2) } \nonumber \\
&& = \Zset_2[\omega] \otimes \Zset_2[{w}_4,\bar{w}_4, {w}_8, \bar{w}_8] \otimes
\Zset_2[{ w}_2, { w}_3, { w}_5, \ldots ]  \label{cSG1B} \\
\lefteqn{H^*(B\Spin_G(1,5), \Zset) } \nonumber \\
&& = \Zset[\varpi] \otimes \Zset[{\newQ}_1,\bar{\newQ}_1,
{\newQ}_2,\bar{\newQ}_2] \otimes \Zset[{W}_3, {W}_5, \ldots] \label{cSG2B} ~.
\ee
As a part of the considerations of Section \ref{Obs}, we identified the degree five generator in the former cohomology ring as a Stiefel-Whitney class. Though the structure of these cohomologies as freely generated rings follows from the analysis, the action of the Steenrod algebra, or the modulo-two-reduction properties of integral generators, does not necessarily follow from those valid for $\Sp_{2}$ or $\Spin(1,5)$.  

The appearance of the second and the  third Stiefel-Whitney
classes there means that the underlying orientable manifold does
not need to be spin or even spin$^{\Cset}$. The corresponding
classes for the normal bundle are absent, effectively because it
was assumed that $\Spin_{G}(1,5)$ was the extension of $\SO(1,5)$
by the spin group of the normal bundle. Note that this is
justified already by the {\em local} structure of the world-volume
theory as a six-dimensional $N=2$ chiral tensor theory.
 
In order to specify such a lift, one has to choose a generalised spin structure. This amounts, effectively, to picking certain classes in the cohomology of the brane. More precisely,  generalised spin structures are classified in 
\be
H^{4}(M, \Zset) \oplus H^{5}(M, \Zset_{2}) ~,  \label{class}
\ee 
in the sense that any two lifts differ by structure that can be characterised fully by an element in (\ref{class}). One may ask whether these classes could be interpreted in terms of fixing characteristic classes of an $\Sp_{2}$ (or an $\SO(5)$) bundle with which the twisting is done. As the structure group of the generalised spin bundle concerns really the semi-direct product of this group and $\Spin(1,5)$, one cannot, in general, translate  characteristic classes of the total $\Spin_{G}(1,5)$ bundle to characteristic classes of the normal bundle.

In the special case where the $\Sp_{2}$ bundle does exist as a lift of a vector bundle, say $NM$, one can give such an identification, though. Indeed, given such a (trivially) generalised spin bundle $\xi \in [M,B\Spin_G(1,5)]$ and $\hat\zeta = p_{6*}\xi$, the degree four part of the generalised spin structure $ \hat\zeta^{*}\bar{\newQ}_{1}$ would correspond  to the half Pontryagin class  $\lambda(NM) = p_{1}(NM)/2$ of the underlying vector bundle. This determines the degree four class $w_{4}(NM)=r(\lambda(NM))$. The remaining piece of generalised spin structure does not seem to be related directly to characteristic classes of the normal bundle, not even in such an hypothetical case as above.

\subsection{Spin structure in the bulk}

Thus far using information   from the bulk has been carefully avoided. It is interesting to note,
nevertheless, that when the 11-dimensional background is spin, one
can define also there the half Pontryagin class $\lambda(TX)$ that
was responsible for shifts in the quantisation condition in the
bulk in \cite{Witten:1996md}. This gives the integral characteristic class
\be 
\lambda(TM) &\eq& i^{*}\lambda(TX) - \hat\zeta^{*}\bar{\newQ}_{1}  ~,
\ee 
even though $M$ does not need to be spin. The terminology is justified as, if the $\Sp_{2}$ and $\Spin(1,5)$ bundles did exist independently, this would be the pertinent half Pontryagin class. 

The existence of such a class would seem to indicate that its reduction modulo two should yield the class $w_{4}(TM)$. This is certainly true when the five-brane is spin. If this is indeed the case, then the topology of such a five-brane embedding is characterised by the constraint \cite{Kalkkinen:2004hs}   
\be
W_{5}(TM) &=& 0 ~,  \label{outo}
\ee
as this is equivalent to $w_{4}(TM)$ being a reduction of an integral class. 

Equation (\ref{outo}) could have arisen  as a standard obstruction
in the discussion of Section \ref{ObsInt}, but did not; it is
rather a consequence of the geometry of the embedding $\iota : M
\hookrightarrow X$, than the existence of global twisted spinors
on the five-brane. The constraint (\ref{outo}) should therefore be
compared rather to the fact \cite{Diaconescu:2000wy} that the
bulk geometry satisfies $W_{7}(TX)=0$, than that the obstruction
to spin$^{\Cset}$ structure would happen to have been $W_{3}(TM)$.

\subsection{Open problems}

In the present case, where the generalised spin group is
specifically fixed to be the physically relevant semidirect
product $\Sp_2 \ltimes \Spin(1,5)$, the vanishing of all
obstructions is automatic (other than the embedding-related
(\ref{outo})). For other extensions $\Spin_{G}(1,5)$ that fit in
(\ref{extension}), the cohomology could be different. In
particular, if the class $W_5$ should then be absent from the
cohomology of $B\Spin_G(1,5)$, it would appear as the obstruction
in (\ref{outo}).

There are many open problems with the dynamics and geometry  of
five-branes, as there are several approaches to describing them.
One of them is in terms of embedded smooth submanifolds, others
include for instance M(atrix) theory constructions
\cite{Banks:1996vh}. The point of view  taken in this paper
extends the class of geometrically described brane configurations
by showing that some structures, such as generalised spin
structure,  make sense on the brane quite irrespective of the
details of how the brane is coupled to the bulk.

It would be interesting to relax further some of the assumptions
made in the beginning, namely that spaces involved should be
orientable. In fact, the bulk M-theory is known to possess a
reflection symmetry \cite{Witten:1996hc,Witten:1995em}; also the
fact that the world-volume theory involves self-dual three-forms
may mean that requiring a Lagrangean formulation in terms of a
local action integral may be too restrictive. Among other matters,
this would lead to more complicated twists in  the cohomology
where the obstructions take there values and, perhaps, provide
insights in five-branes as quantum mechanical solitonic objects in
M-theory.

Finally, it is interesting to note that in their recent work, where they construct partition functions for anti-self-dual tensor theories such as the five-brane world-volume theory, Belov and Moore \cite{talk}  make use of structures that involve choices differing by elements $\mu \in H^{4}_{\text{tors}}(M,\Zset)$. It would be interesting to understand precisely the relationship of that structure to the choice of a generalised spin structure in the present paper.


\subsubsection*{Acknowledgements}
I would like to thank D.~Belov and C.~Isham for discussions. This research is supported by a Particle Physics and Astronomy Research Council (PPARC) Postdoctoral Fellowship.



\begin{thebibliography}{10}

\bibitem{Howe:1983fr}
P.S.~Howe, G.~Sierra and P.K.~Townsend,
``Supersymmetry in Six Dimensions,''
Nucl.\ Phys.\ B {\bf 221} (1983) 331.

\bibitem{Duff:1995wd}
  M.J.~Duff, J.T.~Liu and R.~Minasian,
  ``Eleven-Dimensional Origin of String/String Duality: A One-Loop Test,'' Nucl.\ Phys.\ B {\bf 452} (1995) 261
  [{\tt hep-th/9506126}].

\bibitem{Witten:1996md}
E.~Witten, ``On Flux Quantization in M-theory and the Effective
Action,'' J.\ Geom.\ Phys.\  {\bf 22} (1997) 1 [{\tt
hep-th/9609122}].

\bibitem{Howe:1997fb}
  P.S.~Howe, E.~Sezgin and P.C.~West,
  ``Covariant Field Equations of the M-Theory Five-Brane,''
  Phys.\ Lett.\ B {\bf 399}, 49 (1997)
  [{\tt hep-th/9702008}].

\bibitem{Cremmer:1978km}
E.~Cremmer, B.~Julia and J.~Scherk, ``Supergravity Theory in
Eleven  Dimensions,'' Phys.\ Lett.\ B {\bf 76} (1978) 409.

\bibitem{Avis:1979de}
S.J.~Avis and C.J.~Isham, ``Generalized Spin Structures on Four
Dimensional Space-Times,'' Commun.\ Math.\ Phys.\  {\bf 72}
(1980)103.

\bibitem{Freed:1998tg}
  D.~Freed, J.A.~Harvey, R.~Minasian and G.W.~Moore,
  ``Gravitational Anomaly Cancellation for M-Theory Five-Branes,''
  Adv.\ Theor.\ Math.\ Phys.\  {\bf 2} (1998) 601
  [{\tt hep-th/9803205}].

\bibitem{Maldacena:1997re}
  J.M.~Maldacena,
  ``The Large $N$ Limit of Superconformal Field Theories and Supergravity,''
  Adv.\ Theor.\ Math.\ Phys.\  {\bf 2} (1998) 231
  [Int.\ J.\ Theor.\ Phys.\  {\bf 38} (1999) 1113]
  [{\tt hep-th/9711200}].

\bibitem{Witten:1996hc}
  E.~Witten,
  ``Five-Brane Effective Action in M-Theory,''
  J.\ Geom.\ Phys.\  {\bf 22} (1997) 103
  [{\tt hep-th/9610234}].

\bibitem{Quillen}
D.~Quillen, ``The $\mathrm{mod} ~ 2$ Cohomology Rings of
Extra-Special $2$-Groups and the Spinor Groups,''  Math.\ Ann.\
{\bf 194} (1971) 197.

\bibitem{Lawson}
H.B.~Lawson and  M.-L.~Michelsohn, ``Spin Geometry,'' \\
(Princeton University Press, Princeton, 1989).

\bibitem{Thomas}
E.~Thomas, ``On the Cohomology Group of the Classifying Space for the Stable Spinor Group,'' Bol.\ Soc.\ Mat.\ Mexicana  {\bf 7} (1962) 57.

\bibitem{Borel} A.~Borel, ``Topics in the Homology Theory of Fibre
Bundles," Lect.\ Notes in Math.\ {\bf 36} (1967) 1.

\bibitem{Brown}
E.H.~Brown Jr., ``Cohomology of $B\SO_n$ and $B\O_n$ with Integer
Coefficients,'' Proc.\ Amer.\ Math.\ Soc.\ {\bf 85} (1982) 283.

\bibitem{McCleary}
J.~McCleary, ``A User's Guide to Spectral Sequences,'' (Cambridge
University Press 2001).

\bibitem{Borel2}
A.~Borel, ``Sur l'homologie et la cohomologie des groupes de Lie compacts connexes,''  Amer.\ J.\ Math.\ {\bf 76} (1954) 273.

\bibitem{Robinson}
C.A.~Robinson, ``Moore-Postnikov Systems for Non-Simple
Fibrations,''  Illinois J.\ Math.\  {\bf 16}  (1972) 234.

\bibitem{Kalkkinen:2004hs}
  J.~Kalkkinen,
  ``Holonomies of Intersecting Branes,''
  Fortsch.\ Phys.\  {\bf 53} (2005) 913
   [{\tt hep-th/0412166}].

\bibitem{Diaconescu:2000wy}
  D.E.~Diaconescu, G.W.~Moore and E.~Witten,
  ``$E_{8}$ Gauge Theory, and a Derivation of K-Theory from M-Theory,''
  Adv.\ Theor.\ Math.\ Phys.\  {\bf 6} (2003) 1031
  [{\tt hep-th/0005090}].
  
\bibitem{Banks:1996vh}
  T.~Banks, W.~Fischler, S.~H.~Shenker and L.~Susskind,
  ``M-Theory as a Matrix Model: A Conjecture,''
  Phys.\ Rev.\ D {\bf 55} (1997) 5112
  [{\tt hep-th/9610043}].

\bibitem{Witten:1995em}
  E.~Witten,
  ``Five-Branes and M-Theory on an Orbifold,''
  Nucl.\ Phys.\ B {\bf 463} (1996) 383
  [{\tt hep-th/9512219}].

\bibitem{talk}
  D.~Belov and G.W.~Moore,
  ``Holographic Action for the Self-Dual Field,''
  [{\tt hep-th/0605038}].

\end{thebibliography}
\end{document}